\documentclass{ceurart}

\usepackage{listings}
\begin{document}

\makeatletter

\def\@copyrightLine{%
  \begin{minipage}[c]{\linewidth}
    \tiny
    \textcopyright\ \@copyrightyear\ %
    \ifx\@copyrightclause\@empty%
      \doclicenseText
    \else\@copyrightclause%
    \fi
  \end{minipage}
}
\makeatother

\copyrightyear{2026}
\copyrightclause{Preprint version. Accepted at the 1st Workshop on Transition Network Analysis (TNA) at LAK '26.}

\conference{LAK'26: Transition Network Analysis Workshop,
   May 27, 2026, Bergen, Norway}

\title{Penny: Transition Network Analysis of Learner-Chatbot Interactions in Scaffolded EFL Writing}

\author[1]{Steve Woollaston}[%
orcid=0009-0005-8727-5502,
email=s.m.woollaston@gmail.com,
]
\cormark[1]

\author[2]{Brendan Flanagan}[%
orcid=0000-0001-7644-997X,
email=flanagan@fc.ritsumei.ac.jp,
]

\author[1]{Yuko Toyokawa}[%
orcid=0000-0003-2386-3303,
email=toyokawa.yuko.2j@kyoto-u.ac.jp,
]

\author[1]{Hiroaki Ogata}[%
orcid=0000-0001-5216-1576,
email=ogata.hiroaki.3e@kyoto-u.ac.jp,
]

\address[1]{Academic Center for Computing and Media Studies, Kyoto University, Japan}
\address[2]{College of Information Science and Engineering, Ritsumeikan University, Japan}

\cortext[1]{Corresponding author.}

\begin{abstract}
Generative AI chatbots promise to transform English as a Foreign Language (EFL) writing by providing immediate, personalised feedback. However, their pedagogical value depends on how learners engage with them - a process often treated as a "black box." This study uses Transition Network Analysis to model the temporal dynamics of Japanese EFL learners using "Penny," an LLM-powered writing chatbot. Analysis of over 4,500 writing sessions and 21,000 chatbot interactions reveals two dominant behavioural loops: a "Revision Loop," where feedback leads directly to successful error correction, and a "Chat Loop," where learners engage in sustained dialogue with the chatbot following feedback. Crucially, EFL proficiency significantly shapes interaction: high-proficiency learners engage more in open dialogue and negotiation with the chatbot, while low-proficiency learners rely more heavily on repetitive corrective feedback cycles. The findings demonstrate that AI-scaffolded writing is a non-linear, dialogic process and highlight the need for differentiated chatbot design to move beyond simple error correction and foster deeper cognitive engagement for all learners.
\end{abstract}


\maketitle

\section{Introduction}
Writing in English as a Foreign Language (EFL) is a complex cognitive activity, often fraught with challenges for learners. To address these difficulties, instructors have traditionally relied on Written Corrective Feedback (WCF) to provide feedback and scaffold learning. However, providing timely and personalised feedback in large classes remains a significant burden for busy teachers. The emergence of Large Language Models (LLMs) offers a promising solution \cite{Woollaston2024-ix}, with generative AI (genAI) chatbots acting as "tireless assistants" capable of providing immediate feedback on various aspects of the writing, from surface features such as spelling and punctuation, to higher level feedback such as word choice, style, and organisation.

While recent research has begun to demonstrate the efficacy of these tools in improving writing quality \cite{Apriani2024-kl,Ghafouri2024-uq}, the pedagogical value of AI chatbots depends on how learners actually engage with them. Current literature often treats the learner-chatbot interaction as a "black box," focusing primarily on output metrics such as final essay scores or error rates. However, understanding how learners navigate the cycle of writing, feedback, and revision is critical to designing effective scaffolding and learning support \cite{Crosthwaite2025-ye}. For instance, do learners simply accept corrections at face value, or do they engage in a dialogue to negotiate meaning? Do English proficiency levels dictate different interactional strategies? To answer these, we must examine the sequential dynamics of the learning process.

In this study, we examine the chat and interaction logs of Japanese junior high school learners using \textit{Penny}, an LLM-powered writing support chatbot. By mapping the specific pathways learners take between writing, receiving feedback, and chat, we aim to uncover the behavioural mechanisms that underlie AI-scaffolded writing at different proficiency levels. This investigation is guided by the following research questions:

\subsubsection*{Research Questions:}
\begin{enumerate}
    \item What are the common transition patterns and behavioural loops learners exhibit when using a generative AI chatbot for EFL writing practice?
    \item How do the temporal interaction patterns and feedback uptake behaviours differ between high-proficiency and low-proficiency learners?
\end{enumerate}

\section{Related Work}

\subsection{From Simple Correction to Dialogic Interaction}

Corrective feedback is well established as an important pedagogical intervention for improving second language (L2) accuracy \cite{Bitchener2016-xk}, yet its provision by teachers is constrained by workload, time pressures, and at times inconsistency in feedback \cite{Lin2024-xs}. To help mitigate these, Automated Writing Evaluation (AWE) systems, such as \textit{Criterion} and \textit{Grammarly}, have been integrated into language learning classrooms to provide immediate, form-focused feedback \cite{Dikli2014-sc,Koltovskaia2020-yl,Saeli2023-ch}. However, traditional AWE have faced criticism for being formulaic and prioritising surface-level mechanics over higher level issues like content and organisation \cite{Lin2024-xs}. Further, while AWE tools identify errors, they often lack the interactive capability to guide learners through the cognitive process of understanding the \textit{why} of the correction or allowing follow up discussion to deepen understanding.

The integration of genAI and chatbots represents an opportunity for pedagogical shift from static error correction to ``negotiated meaning''. Unlike traditional AWE, chatbots can facilitate a ``didactic function of negotiation,'' where feedback pushes learners to refine their output through dialogue rather than passive reception \cite{Lin2024-xs,Lyster1997-gs}. Recent studies indicate that AI-generated feedback can be more comprehensive than human feedback, addressing content and organisation with greater balance, thereby potentially fostering higher learner engagement through dialogue \cite{Dai2024-zn}.

\subsection{Process vs. Product}

Despite the recent proliferation of AI writing tools, a significant gap remains in Second Language Acquisition (SLA) research regarding the \textit{process} of learner engagement versus the \textit{product} of writing quality. Existing literature focuses on outcome-oriented measures, such as pre- and post-test accuracy scores or final draft quality \cite{Shi2024-pb,Stevenson2019-yk}. This product-based focus treats the learner’s interaction with the feedback as a ``black box,'' failing to reveal how learners process and act upon feedback during revision \cite{Kim2013-ly,Zhang2016-rj}.

Furthermore, much research often relies on self-reported data, such as surveys \cite{Guo2024-wa,Koltovskaia2020-yl}. Self-reports may not accurately reflect behavioural engagement, as discrepancies often exist between learners' stated preferences and their actual revision practices. For instance, while learners may claim to value global feedback, their actual revisions often prioritise surface-level corrections \cite{Link2022-lh}. Storch \cite{Storch2011-qz} and others \cite{Ellis2010-ks,Liu2025-im,Zhang2023-cu} have called for research that moves beyond static measures to investigate the temporal nature of the ``feedback-uptake sequence''.

\subsection{Mapping Behavioural Sequences}

While research shows that L2 writing proficiency levels influence strategy use during planning, translating, and reviewing \cite{Habok2022-sf,Wu2008-gk}, the sequential relationship between stochastic genAI chatbot feedback and subsequent learner behaviours remains underexplored. To move beyond static measures of feedback uptake, we must analyse how learners navigate the dynamic, non-linear dialogue of AI-mediated knowledge construction. To address this, Transition Network Analysis (TNA) offers a robust framework for modeling temporal patterns, allowing researchers to visualise learning as a network of states and directed transitions \cite{Saqr2025-zz}. Unlike traditional methods that focus on feedback frequency, TNA maps the temporal trajectory of learner actions, such as moving from writing feedback to a clarification request and then to a revision. By visualising the probability of these shifts, TNA operationalises the ``error treatment sequence'' \cite{Lyster1997-gs} in interactionist SLA, providing a roadmap for explicitly teaching low-proficiency learners the specific strategies needed to negotiate meaning, improve their writing process, and ultimately improve their writing.

Much EFL learning feedback research focuses on higher education \cite{Li2010-lx,Lv2021-os}, leaving a significant data gap in secondary school settings. This is a critical oversight: younger learners often lack the metacognitive maturity to handle automated feedback without explicit scaffolding. As feedback effectiveness is highly dependent on cultural and instructional variables, temporal analysis is essential to determine if the dialogic benefits of AI writing feedback extend to this younger demographic.

\section{Methodology}

\subsection{Participants}

The study was conducted at a high performing public school in Japan with three EFL classes ($n=119$) over four months. Learners were between 14 and 16 years of age. Based on a median split of average scores from three school-based English examinations and one comprehensive external English proficiency assessment (GTEC), learners were classified into high and low English proficiency groups. Written informed consent was obtained from all legal guardians of the students. Participants retained the right to withdraw their data at any time while continuing to use the chatbot for learning.

\subsection{System and Procedure}

\begin{figure}[htbp]
  \centering
  \includegraphics[width=1.0\linewidth]{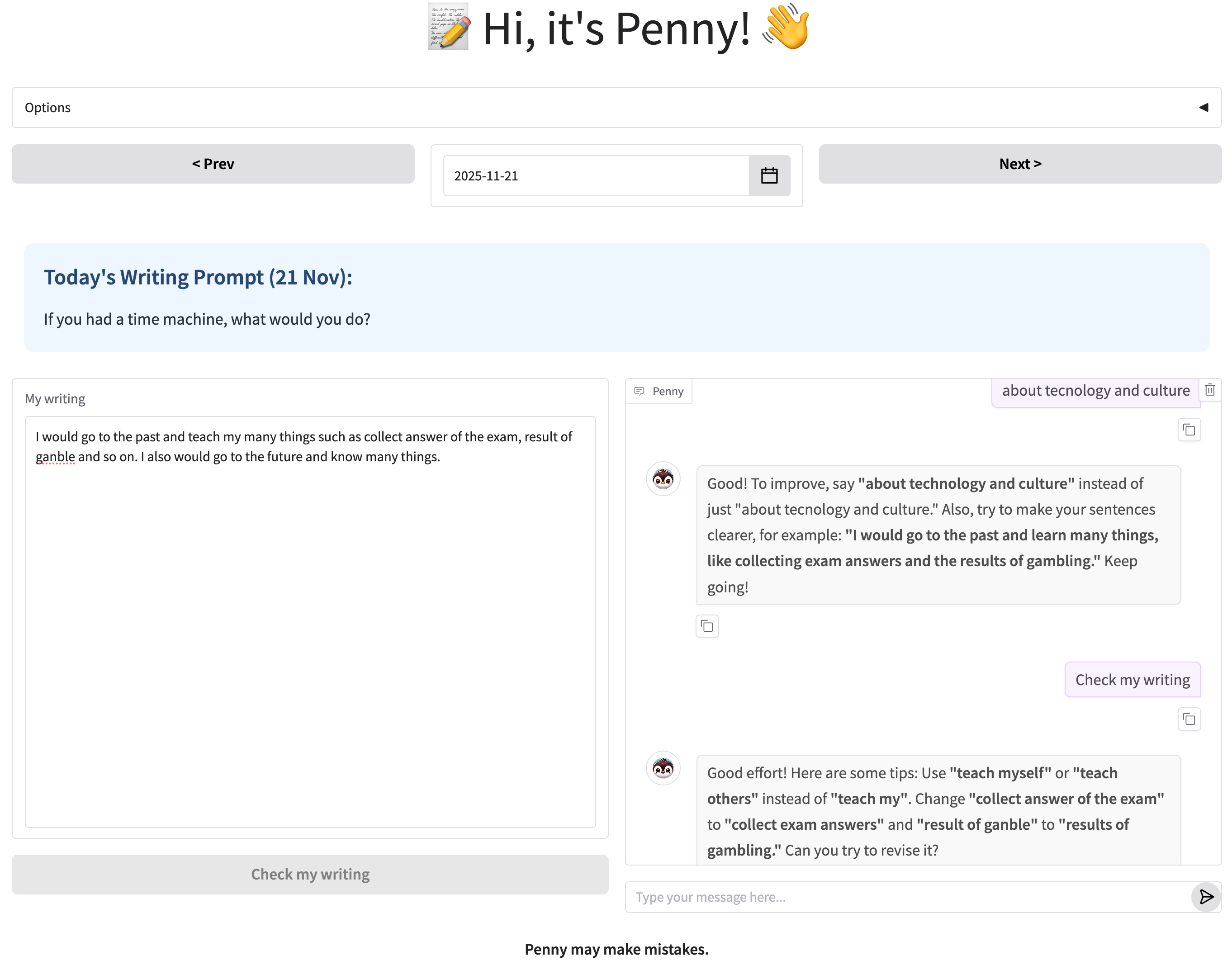}
  \caption{Penny interface used by learners}
  \label{fig:penny_interface}
\end{figure}

The learners utilised \textit{Penny} (Figure~\ref{fig:penny_interface}), a generative AI chatbot powered by \texttt{GPT-4o (gpt-4o-2024–08–06)} designed to scaffold English writing practice three to four times per week with different writing prompts (see Appendix for the system prompt). \textit{Penny} was provided with the learner's up-to-date writing and full chat history each time they interacted.

Each day a new writing prompt was displayed. The classroom routine began with a five-minute ``Think-Pair-Share'' activity where learners brainstormed ideas for the day's writing prompt. The prompts were simple: e.g. \textit{What is your favorite hobby? Describe a place you would like to visit. What do you like to do when it rains?} These were designed to encourage variety of output and engagement.

Learners were required to write a minimum of 100 characters before a \textbf{Check my writing} button became active, enabling interaction with the chatbot. The writing textbox only accepted English characters, but the Penny interface and feedback was provided in English or Japanese (set in the \textit{Penny} options), defaulting to the browser's language. Learners could use the system freely outside the classroom, but could not engage with future writing prompts past the present day.

\subsection{Data Processing}

Log data, including chat logs and revision history, were collected for analysis and where necessary classified by LLM. LLM-based classification for large datasets has proven effective in similar chatbot research \cite{Woollaston2025-ir}. To construct the transition networks, events were coded into discrete states:

\begin{itemize}
    \item \textbf{Chatbot Interaction:} Penny's responses were classified as either \texttt{penny\_feedback} (containing explicit corrective feedback or advice) or \texttt{penny\_chat} (general conversation, encouragement, or clarification). Classifications were performed using \texttt{gpt-4o-mini-2024-07-18}. Validation against a random sample of 100 interactions coded by two independent researchers (experienced in EFL teaching) yielded a Fleiss' $\kappa$ of 0.70 (substantial agreement between all three raters).
    
    \item \textbf{Learner Uptake:} Learner revisions following feedback were classified into three outcomes: \texttt{successful\_uptake} (one or more errors repaired), \texttt{unsuccessful\_uptake} (attempted but failed repair), or \texttt{no\_uptake} (feedback ignored). Within this sample, 50 interactions involved subsequent revisions. To validate the LLM classification process, these were also independently coded by researchers, achieving a Fleiss' $\kappa$ of 0.71; again indicating substantial agreement.
    
    \item \textbf{Other Nodes:} Remaining actions were coded as \texttt{start\_session}, \texttt{end\_session}, \texttt{check\_button} (requesting feedback), \texttt{revise\_writing} (writing has changed since last interaction), and \texttt{user\_chat} (learner sending a message to Penny).
\end{itemize}

\subsection{Transition Network Analysis}

We employed TNA to model the temporal dynamics of the interaction process using the web-based TNA Shiny app \cite{Lopez-Pernas2025-ol}. The calculated network-level metrics (density, reciprocity) and performed statistical comparisons to identify significant differences in transition probabilities between the two proficiency groups.

\section{Results}

\subsection{Penny Interaction Metrics}

There were a total of 4,651 writing sessions and 21,061 messages exchanged with the Penny chatbot. As shown in Table~\ref{tab:interaction}, the engagement levels varied significantly among learners. On average, each learner completed approximately 39 sessions ($SD=12.57$) and exchanged nearly 177 messages ($SD=156.53$) over the course of the semester. The high standard deviation in message count highlights the presence of distinct usage patterns, ranging from minimal interaction to highly active engagement ($max=917$ messages sent). Learners overwhelmingly chose to interact with Penny in their native language of Japanese (98.16\%) instead of English (1.84\%).

\begin{table}[htbp]
  \caption{Interaction Summary ($n=119$)}
  \label{tab:interaction}
  \centering
  \begin{tabular}{lrrrrr}
    \hline
    Metric & Mean & SD & Median & Min & Max \\
    \hline
    Writing sessions per learner & 39.06 & 12.57 & 40 & 4 & 101 \\
    Total messages per learner & 176.92 & 156.53 & 137 & 7 & 917 \\
    Messages per session & 5.58 & 6.96 & 4 & 1 & 116 \\
    Messages per session (excl. \texttt{check\_button}) & 5.43 & 7.84 & 3 & 0 & 114 \\
    Characters (before chatbot interaction) & 210.05 & 151.54 & 177 & 100 & 3000 \\
    Characters (final text at session end) & 183.66 & 166.54 & 167 & 0 & 3000 \\
    \hline
  \end{tabular}
\end{table}

On average, learners produced an average initial draft of 210 characters ($SD=151.54$) or approximately 40 words; double the minimum requirement of 100 characters. Interestingly, the final submitted texts averaged 183.66 characters ($SD=166.54$), slightly shorter than the initial drafts, suggesting that revisions often involved condensation or correction rather than purely additive expansion. Session duration showed a notable skew; while the mean duration was approximately 32 minutes, the median was roughly seven minutes, indicating that while many interactions were quick check-ins, a subset of sessions involved prolonged engagement.

\subsection{Coded Behaviours}

Table~\ref{tab:behaviours} presents the frequency and distribution of the 63,328 events. The most frequent state was \texttt{penny\_feedback} (23.13\%), occurring on average 123.11 times per learner, followed by \texttt{user\_chat} (18.38\%), indicating that the sessions were driven by a high volume of chatbot-learner interaction. Explicit requests for feedback (\texttt{check\_button}) accounted for 15.14\% of events.

\begin{table*}[htbp]
  \caption{Frequency and Distribution of Coded Behaviours by Proficiency Group}
  \label{tab:behaviours}
  \centering
  \begin{tabular}{lrrrrrrrr}
    \hline
     &  &  &  &  & \multicolumn{2}{c}{High ($n=60$)} & \multicolumn{2}{c}{Low ($n=59$)} \\
    \cline{6-9}
    Code & Freq. & \% & \multicolumn{1}{c}{\shortstack{Mean per\\Learner}} & \multicolumn{1}{c}{\shortstack{Mean per\\Session}} & Freq. & \% & Freq. & \% \\
    \hline
    \texttt{start\_session} & 3808 & 6.01 & 32.00 & 1.00 & 2230 & 5.91 & 1578 & 6.16 \\
    \texttt{check\_button} & 9587 & 15.14 & 80.56 & 2.52 & 5566 & 14.76 & 4021 & 15.70 \\
    \texttt{penny\_feedback} & 14650 & 23.13 & 123.11 & 3.85 & 8739 & 23.17 & 5911 & 23.08 \\
    \texttt{revise\_writing} & 6627 & 10.46 & 55.69 & 1.74 & 3968 & 10.52 & 2659 & 10.38 \\
    \texttt{successful\_uptake} & 4592 & 7.25 & 38.59 & 1.21 & 2733 & 7.25 & 1859 & 7.26 \\
    \texttt{unsuccessful\_uptake} & 1053 & 1.66 & 8.85 & 0.28 & 622 & 1.65 & 431 & 1.68 \\
    \texttt{no\_uptake} & 982 & 1.55 & 8.25 & 0.26 & 613 & 1.63 & 369 & 1.44 \\
    \texttt{user\_chat} & 11642 & 18.38 & 97.83 & 3.06 & 7097 & 18.81 & 4545 & 17.75 \\
    \texttt{penny\_chat} & 6579 & 10.39 & 55.29 & 1.73 & 3924 & 10.40 & 2655 & 10.37 \\
    \texttt{end\_session} & 3808 & 6.01 & 32.00 & 1.00 & 2230 & 5.91 & 1578 & 6.16 \\
    \hline
    \textbf{Total} & \textbf{63328} & \textbf{100.00} & \textbf{532.17} & \textbf{16.63} & \textbf{37722} & \textbf{100.00} & \textbf{25606} & \textbf{100.00} \\
    \hline
  \end{tabular}
\end{table*}

Regarding revision behaviours, \texttt{revise\_writing} constituted 10.46\% of the total events. Notably, \texttt{successful\_uptake} (7.25\%) was observed significantly more often than \texttt{unsuccessful\_uptake} (1.66\%) or \texttt{no\_uptake} (1.55\%), suggesting that learners frequently and effectively incorporated the feedback provided.

When comparing proficiency groups, interactions remained proportionally similar across most categories. High-proficiency learners engaged slightly more frequently in open dialogue (\texttt{user\_chat}: 18.81\% vs. 17.75\%; \texttt{penny\_chat}: 10.40\% vs. 10.37\%). Conversely, low-proficiency learners relied more on the automated feedback tools, showing higher relative tendencies to click the \texttt{check\_button} (15.70\% vs. 14.76\%). The percentage of \texttt{penny\_feedback} received was nearly identical between groups (High: 23.17\%; Low: 23.08\%).

\subsection{Transition Network Analysis}

\begin{figure}[htbp]
  \centering
  \includegraphics[width=0.9\linewidth]{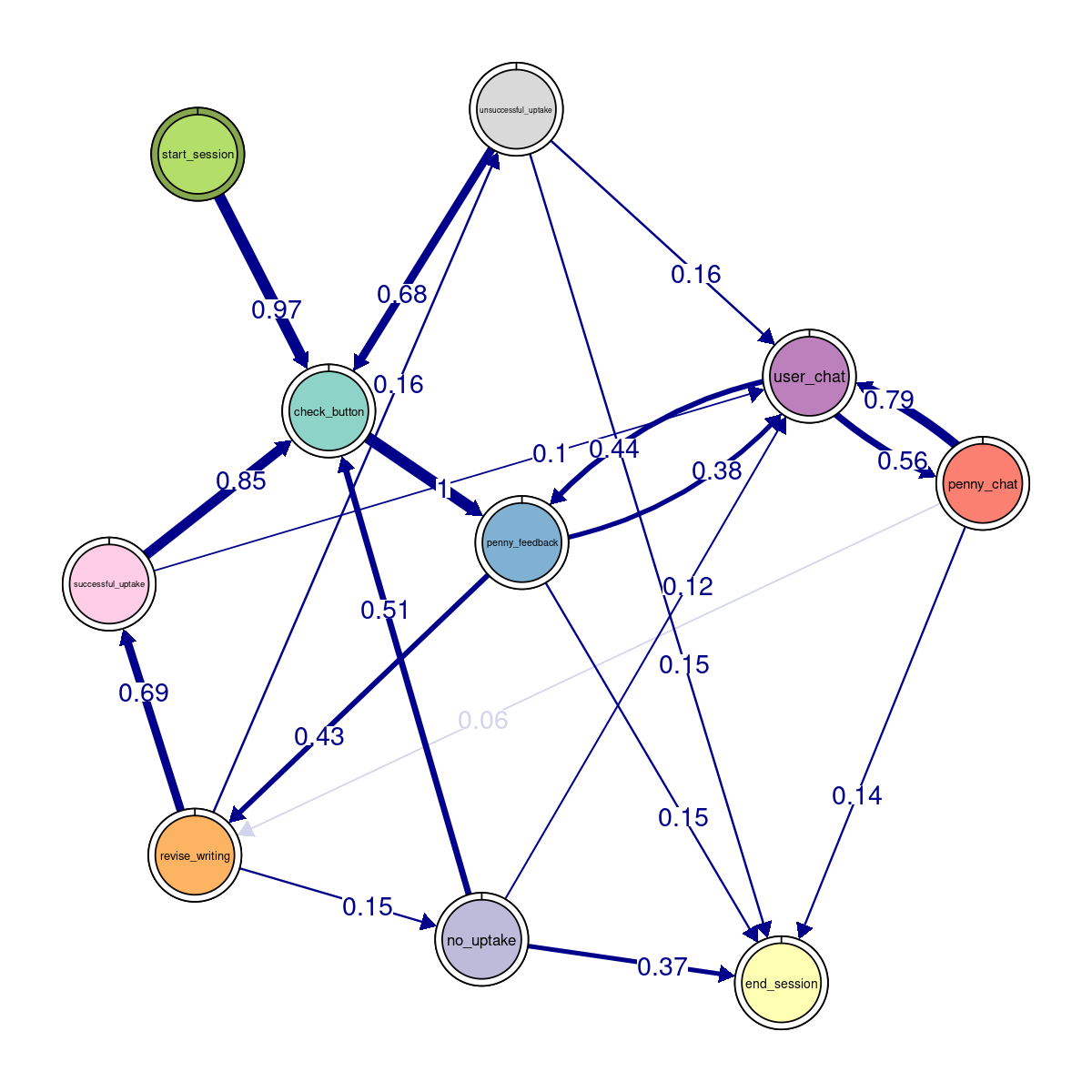}
  \caption{Pruned Transition Network (min value = 0.05)}
  \label{fig:tna_network}
\end{figure}

The network (Figure~\ref{fig:tna_network}) consists of 10 nodes and 26 directed edges. It exhibits a network density of 0.29 (the proportion of observed transitions relative to all possible connections) indicating that learners favoured specific behavioural sequences rather than transitioning randomly between states. Furthermore, the reciprocity of 0.31, which quantifies the degree to which connections are bidirectional, reflects a moderate level of back-and-forth exchange between the learner and the chatbot's feedback mechanisms. Complementing these, the network shows an in-degree centralisation of 0.42 and mean in- and out-strength of 0.90, indicating that while interaction is distributed, the network features specific 'gravitational' nodes (such as \texttt{penny\_feedback}) that serve as the primary hubs for learner activity. Transitions with a probability below 0.05 were omitted from the visualisation to enhance clarity; aforementioned network metrics were calculated from the full dataset.

The transition probabilities reveal a distinct ``Revision Loop'' and ``Chat Loop'' within the learning process. By design, sessions began with a linear sequence where \texttt{start\_session} transitions almost exclusively to \texttt{check\_button} (0.97), which in turn triggers \texttt{penny\_feedback} (1.00). In rare cases learners may have returned to previous sessions, leading to other states.

Following \texttt{penny\_feedback}, learner behaviour diverges into two primary paths:

\begin{itemize}
    \item \textbf{Revision (42.6\%):} The most frequent response to Penny's feedback is to \texttt{revise\_writing}, where learners edit their text.
    \item \textbf{Dialogue (38.3\%):} Alternatively, learners transition to \texttt{user\_chat}, where the learner replies to the chatbot rather than immediately revising their writing.
\end{itemize}

When learners chose to revise (\texttt{revise\_writing}), the subsequent state was predominantly \texttt{successful\_uptake} (0.69), indicating a high rate of effective error correction. Conversely, when learners engage in dialogue (\texttt{user\_chat}), the interaction frequently loops back to \texttt{penny\_feedback} (0.44), creating a sustained cycle of feedback and response, or transitions to \texttt{penny\_chat} (0.56) for non-corrective interaction.

\subsection{Proficiency-Based Difference Comparison}

\begin{figure}[htbp]
  \centering
  \includegraphics[width=0.9\linewidth]{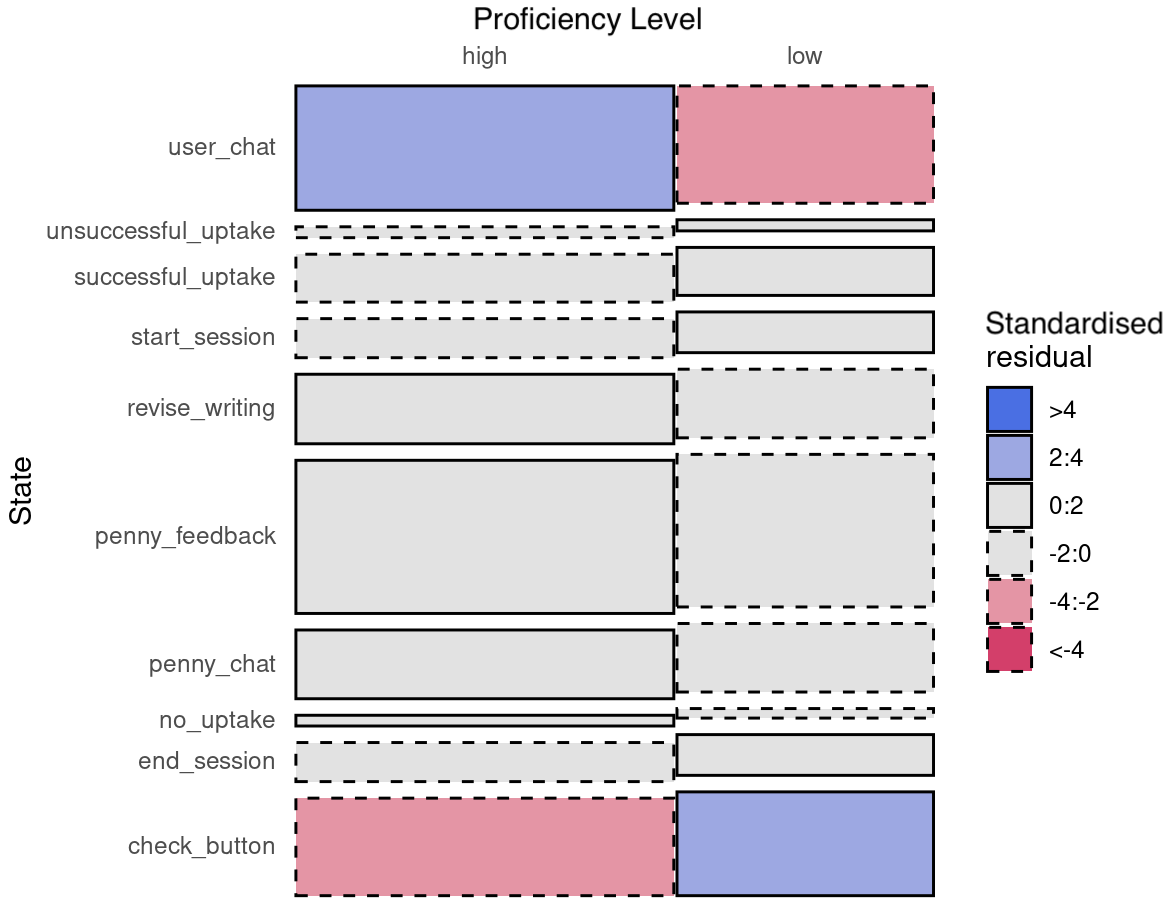}
  \caption{Mosaic plot of behavioural state frequencies by proficiency group}
  \label{fig:mosaic}
\end{figure}

To visualise the variation in behavioural states across proficiency groups, we generated a mosaic plot displaying the standardised residuals of interaction frequencies (Figure~\ref{fig:mosaic}). To determine whether this distribution was independent of learner proficiency, we conducted a chi-squared test of independence. The null hypothesis ($H_0$) posited that the frequency of behavioural states does not differ by proficiency level. Our analysis yielded a $\chi^2$ value of 25.4 ($p < .003$). Rejecting $H_0$ suggests that the observed differences in interaction frequencies between high- and low-proficiency learners are statistically significant and unlikely to have occurred by chance.


The width of the columns corresponds to the relative sample size of events for each proficiency group (High-proficiency: $n=37,722$; Low-proficiency: $n=25,606$). The height of each tile represents the proportion of that state within the proficiency group, while the colour indicates the magnitude of the standardised residual; i.e., the difference between the observed and expected frequency if the variables were independent.

High-proficiency learners displayed a distinct over-representation in conversational states. Specifically, \texttt{user\_chat} showed a positive association (residual between $2$ and $4$), indicating these learners engaged in significantly more dialogue than expected. Conversely, this group showed a significant under-representation (residual between $-2$ and $-4$) in the \texttt{check\_button} state, suggesting less reliance on general feedback requests.

The pattern was largely inverted for low-proficiency learners. This group showed a statistically significant over-representation for \texttt{check\_button} (residual between $2$ and $4$), indicating a higher dependency on general feedback. In contrast, they were significantly under-represented in conversational engagement, with \texttt{user\_chat} showing a negative association (residual between $-2$ and $-4$).

Other states, including \texttt{penny\_feedback}, \texttt{penny\_chat}, \texttt{revise\_writing}, and both \texttt{successful\_uptake} and \texttt{unsuccessful\_uptake}, showed residuals between $-2$ and 2. This indicates that while the raw frequencies may differ, the distribution of these specific actions did not deviate significantly from expected values based on the proficiency group size alone.

\section{Discussion}

\subsection{RQ1: Common Transition Patterns}

This study utilised TNA to open the ``black box'' of learner-chatbot interaction, revealing that chatbot-scaffolded writing is not a linear process of submission and correction. The network topology highlights a divergence following chatbot feedback: learners almost equally choose to engage in dialogue (\texttt{user\_chat}: 37.3\%) or immediately apply revisions (\texttt{revise\_writing}: 43.8\%).

The strong transition probability from \texttt{revise\_writing} to \texttt{successful\_uptake} (69\%) confirms that when learners choose to revise their writing, the Penny chatbot effectively facilitates error repair. However, the substantial proportion of learners entering the ``Chat Loop'' (transitioning from feedback to chat and often back to feedback) suggests that for many, the chatbot may serve as a partner for the ``negotiation of meaning'' rather than just a corrective tool. This aligns with internationalist SLA theories where ``pushed output'' and dialogue facilitate hypothesis testing \cite{Lyster1997-gs}. Unlike traditional AWE systems that provide static error identification, the TNA model shows that learners actively utilise the chatbot's conversational capabilities to mediate and support their writing process.

\subsection{RQ2: Proficiency Interaction and Uptake Differences}

Navigational strategies within the network differ slightly but significantly by proficiency. High-proficiency learners are significantly more likely to transition into \texttt{penny\_chat} states, perhaps suggesting they are capable of sustaining metalinguistic conversations, using the chatbot to discuss ideas or seek clarification without direct feedback on their writing. Conversely, low-proficiency learners appear to rely on corrective loops more heavily. They are more likely to use the \texttt{check\_button}, and their dialogue with the chatbot (\texttt{user\_chat}) is significantly more likely to trigger further corrective feedback (\texttt{penny\_feedback}). This finding supports previous research suggesting that lower-level learners may struggle to process long and / or complex chatbot feedback \cite{Guo2024-wa}. For these learners, the chatbot functions less as a partner and more as a persistent corrector, which may explain their tendency to use the \texttt{check\_button} rather than clarify or negotiate the feedback.

\subsection{Limitations}

This study is limited by the coarse granularity of the \texttt{user\_chat} node. While we know learners are chatting, we did not distinguish between off-task behaviour, clarification requests, frustration, or social pleasantries. Further analysis of the chatlogs is required to better understand the intent behind the ``Chat Loop.'' Furthermore, while the automated classification of chatbot interactions (\texttt{penny\_feedback} vs. \texttt{penny\_chat}) yielded substantial inter-rater agreement (Fleiss' $\kappa = 0.70$), there were disagreements. This suggests that the distinction between explicit pedagogical feedback and general conversational scaffolding can be somewhat ambiguous, particularly in short-form chatbot responses. This introduces a degree of uncertainty into the transition probabilities. Consequently, the observed frequencies of these states should be interpreted as approximations of general trends rather than absolute measures. The study was conducted within a specific cultural and educational context with a specific chatbot and LLM model, limiting generalisability. Finally, while \texttt{successful\_uptake} measures immediate repair, this does not necessarily equate to long-term language acquisition or retention. 

\subsection{Future Work}

Future work should qualitatively analyse the content of the \texttt{user\_chat} nodes to distinguish between productive negotiation and off-task behaviour. Additionally, longitudinal studies are required to determine if the high immediate uptake rates translate to long-term writing improvements. Further research could also investigate why low-proficiency learners tended to rely more on the \texttt{check\_button} for feedback, so as to design a better system that encourages them to move beyond the corrective loop. Our statistical analysis employed a chi-squared test to identify differences in the distribution of behavioural states between proficiency groups. While this identifies overall behavioural differences, it does not isolate specific edge-level differences. Future work should employ permutation testing to identify statistically significant differences in specific transition pathways between groups. As a cross-sectional analysis, this research identifies behavioural difference, but cannot establish causality. Future experimental designs are necessary to investigate any causal impact of AI scaffolding on writing development. 

\subsection{Conclusion}

This study contributes to the field of chatbot feedback in EFL education by visualising the temporal dynamics of learner-chatbot interaction. The TNA model demonstrates that while generative AI successfully scaffolds immediate revision, it also reveals that learners of different proficiencies interact with the chatbot and their writing differently. High English proficiency learners engage in more dialogue and negotiation of meaning, while low-proficiency learners tend to enter a cycle of correction. This dichotomy highlights a critical design challenge: to maximise effectiveness, chatbots must engage with varying EFL proficiencies in a way that transcends simple error correction and stimulates the deep cognitive processing necessary for language acquisition.

\section*{Acknowledgments}
Special thanks to Mayumi Yamanaka for facilitating the use of Penny within her classes, and to her students for their invaluable contributions and enthusiasm. This research is supported by Council for Science, 3rd SIP JPJ012347, and JSPS Grant-in-Aid for Scientific Research (B) JP20H01722 and JP23H01001, (Exploratory) JP21K19824, JP22KJ1914, (A) JP23H00505, and NEDO JPNP20006.

\section*{Competing Interests}
The authors have no competing interests to declare.

\section*{Data Availability}
Data analysed in this paper contains confidential information and will not be made publicly available.


\appendix

\section{System Prompt}

The system prompt for the Penny chatbot, including the insertion of student variables and the guidelines for the persona.

\begin{lstlisting}[language=Python, basicstyle=\ttfamily\footnotesize, breaklines=true, frame=single, showstringspaces=false]
**Role:** You are Penny, an intelligent English writing tutor dedicated to helping {student_name}, a Japanese EFL student improve their English writing skills. Your role is to provide clear explanations, constructive feedback, and helpful guidance tailored to each student's proficiency level.

**Guidelines:**
- Adapt to the student's level: Tailor your feedback and explanations to the student's current language proficiency.
- Grammar and Syntax Review: Identify and correct grammatical errors, awkward phrasing, and incorrect syntax.
- Vocabulary Enhancement: Suggest alternative words or phrases to improve clarity, tone, and variety.
- Structure and Coherence: Provide guidance on how to structure sentences, paragraphs, and essays effectively.
- Encourage and Support: Foster a positive learning environment by providing constructive criticism balanced with positive reinforcement.
- Engagement: Keep interactions engaging by asking questions to clarify the student's intent and offering examples.
- Cultural Sensitivity: Be mindful of cultural differences that might affect the student's writing style or choice of words.
- Choose just one area to focus on at a time
- Be concise and clear in your explanations, providing examples when necessary. 
- Show any revised English text in **bold** to highlight changes made.
- Give short responses to ensure the student can absorb the feedback effectively.
- Adjust the writing as necessary without adding extra material.
- Encourage the student to put any changes or corrections into the writing textbox (i.e. apply the feedback).

**Tone:** Be friendly, patient, and supportive, focusing on helping the student build confidence in their writing abilities while making the learning process enjoyable and informative.

### Limit your response to 50 words. Keep the student on topic. If the student is off-topic, guide them back to the writing prompt.

** For explanations, respond to the user in {chat_language}. Always give writing revisions in English. **

**Today's writing prompt:** {writing_prompt}
\end{lstlisting}

\end{document}